\documentstyle[aps,prb,preprint]{revtex}
\begin{document}
\draft
\title{Pair description of the FQHE with
application to edge waves}
\author{M. Marsili\\
{\em International School for Advanced Studies (SISSA)\\
V.Beirut 2-4, 34014 Trieste, Italy}}
\maketitle
\begin{abstract}
Pairs of particles of definite
total and relative angular momentum provide a natural
description for a two dimensional electron gas in a
strong magnetic field. Two body operators take
a simple form when expressed in terms of pair
creation and destruction operators. The pair formalism is
applied to the study of edge waves excitations. For
$\nu=1$ the operators which create edge excitations
are identified and the role the interaction potential
plays in the long wavelength limit is clarified.
This picture is claimed to describe also edge
excitations on the $\nu=1/m$ Laughlin states.
\end{abstract}
\narrowtext

\section{Introduction}

Contrary to the integer quantum Hall effect, which
can be accounted for by a single particle description,
the fractional effect arises as a result of
condensation into a macroscopic collective ground
state~\cite{prang}.
Much of the present understanding of the fractional
quantum Hall effect (FQHE) is based on first quantized
many electrons wave functions~\cite{jain,trial}. The
strategy based on trial "variational" wave functions
has the advantage of displaying in a very direct way
the many body correlations between electrons.
These correlations are induced by the pair potential
acting on the electrons and this naturally leads to
consider pair of electrons as the relevant degrees
of freedom. Indeed the notion of particle pairs has
been used by many authors
\cite{trial,laugh,pseudi,trug,mcdon} in the study
of the FQH effect. No
systematic description of the system in terms of pairs
and no detailed analysis of the the distribution of
their angular momenta has been given up to now.
The first part of this work is an attempt in this
direction. In the symmetric gauge pair
creation and annihilation operators are introduced
in a second quantized formalism; their quantum numbers
are the total angular momentum (TAM) and the relative
angular momentum (RAM).
All relevant two body operators in the lowest Landau level
(LLL) are simply related to the distribution of pairs'
TAM and RAM. In spite of the
simple expression of the operators, the complex
commutation relations among pairs, which deviate from
perfect boson character, preclude a simple description
of the Hilbert space. The first section is
devoted to the derivation of the basic formulas, to the
discussion of some simple examples and to a brief
review of some well known results, easily recovered
within the pair description.

The second part deals with an application to a specific
problem for which an interesting picture has been
recently proposed on the basis of both analytical and
numerical work~\cite{stone}. This concerns the low lying
excitations near incompressible states as the $\nu=1$ or
$\nu=1/3$ states usually called edge waves.
The pair picture provides a simple description of
edge excitations. The main result is an operator relation
that, under some conditions for the incompressible ground
state, allows to identify the creation operators of edge
excitations. The case $\nu=1$, for which these conditions
are satisfied, is treated in detail also relying on a
Hartree approximation. The same picture is
suggested to hold also for the $\nu=1/m$ Laughlin states.
The role the e$^-$- e$^-$ interaction plays in the edge
wave dispersion relation is also clarified. In particular
the results confirm the validity of the semiclassical
approximation\cite{wen} for edge states and suggest that
the dispersion relation is asymptotically linear.
Moreover it is shown that the contribution to the
velocity of edge waves vanishes, in the thermodynamic
limit, for any potential that decays faster
than $1/r$ as $r\to\infty$.

\section{Pair description of the FQHE in the
symmetric gauge}

The Hilbert space of two dimensional charged particles
in the $x$-$y$ plane is split into Landau levels by a
magnetic field $\vec{B}=B_z\hat{z}$. In the extreme
quantum limit ($B_z\rightarrow\infty$) all the particles
are confined in the lowest level. The kinetic energy,
reduced to the zero point motion, is an inessential
constant so that the hamiltonian of the system contains
only potential terms. Among these the dominant role is
played by the particle-particle interaction potential.
With this general motivation in mind we will concentrate
in this section on the projection of two body operators
on the LLL.

Let us consider a general two body operator represented,
in first quantization, by the function ${\cal V}(z_1,z_2)$
($z_j=x_j+iy_j$ is the complex coordinate for particle
$j$ on the complex plane). Magnetic units, $\ell=\sqrt{\hbar
c/eH}=1$, will be used throughout.
The second quantized form of
this operator in the LLL, in the symmetric gauge, is:
\widetext
\begin{equation}
\hat{V}=\frac{1}{2}\sum_{T=1}^\infty \sum_{u,s=0}^T
\langle T-u,u\vert {\cal V}\vert T-s,s\rangle
c^+_{T-u} c^+_u c^{}_s c^{}_{T-s}
\label{one}
\end{equation}
The operator $c^+_u$ creates one electron in the LLL orbital
$\phi_u(z)=\left(2\pi 2^u u!\right)^{-1/2}z^u e^{-|z|^2/4}$
and $c_u$ is its hermitian conjugate. The conservation of
TAM is explicit in equation
(\ref{one}) and the matrix element is given by:
\begin{equation}
\langle T-u,u\vert {\cal V}\vert T-s,s\rangle =
\frac{I_V(T,s,u)}{4\pi^2 2^T\sqrt{(T-s)! s! (T-u)! u!}}
\end{equation}
with
$I_V(T,s,u)= \int\! d^2 z_1 d^2 z_2\, \bar{z_1}^{T-u}
\bar{z_2}^u {\cal V}(z_1,z_2) z_1^{T-s}z_2^s
e^{-(|z_1|^2+|z_2|^2)/2}$.
We consider in what follows only operators
${\cal V}(z_1,z_2)$ that are separable in
the relative ($\xi=(z_1-z_2)/2$) and center of mass
($Z=(z_1+z_2)/2$) coordinates:
${\cal V}(Z+\xi,Z-\xi)=w(Z)\cdot v(\xi)$
In this case the integral in the matrix element can
be performed~\cite{girv} as follows:
\begin{eqnarray*}
I_V(T,s,u) & = &
\sum_{\alpha=0}^{T-s}\sum_{\beta=0}^s\sum_{\gamma=0}^{T-u}
\sum_{\sigma=0}^u{T-s \choose \alpha} {s \choose \beta}
{T-u \choose \gamma} {u \choose \sigma}
(-1)^{\beta+\sigma} \cdot\\
&  & \cdot\int\! d^2 Z\,
\bar{Z}^{T-\gamma-\sigma}w(Z) Z^{T-\alpha-\beta}
e^{-|Z|^2}\cdot 2\int\!d^2\xi\, bar{\xi}^{\gamma+\sigma}
v(\xi)\xi^{\alpha+\beta}e^{-|\xi|^2}.
\end{eqnarray*}
The integrals vanish
whenever $\alpha+\beta\neq \gamma+\sigma$.
Using a differential representation of the discrete
delta function $\delta_{i,j}=\partial_x^i x^j/i!\vert_{x=0}$
and introducing the integer variable
$q=\alpha+\beta =\gamma+\sigma$
the sums on $\alpha$, $\beta$, $\gamma$ and
$\sigma$ can be carried out and
\begin{eqnarray*}
I_V(T,s,u) & = & \sum_{q=0}^T
\left.\frac{\partial_x^q}{q!}(1+x)^{T-u}(1-x)^u \right|_{x=0}
\left.\frac{\partial_y^q}{q!}(1+y)^{T-s}(1-y)^s \right|_{y=0}
 I_\xi(q) I_Z(T,q)\\
\hbox{with} & & I_\xi(q)=2\int \! d^2 \xi\, v(\xi)
|\xi|^{2q} e^{-|\xi|^2} \hbox{,~~}I_Z(T,q)=\int\!
d^2 Z\, w(Z) |Z|^{2(T-q)} e^{-|Z|^2}
\end{eqnarray*}
\narrowtext
It is now possible to sum independently on $u$ and $s$
in equation (\ref{one}), so that the second
quantized form of the operator $\hat{V}$ becomes:
\begin{equation}
\hat{V}=\sum_{T=1}^\infty \sum_{q=0}^T
V_T(q) f^+_T(q)f^{}_T(q).
\label{vop}
\end{equation}

The sum on $u$ in equation (\ref{one}), which involves
the first integral and the pair of operators
$c^+_{T-u}c^+_u$, defines the pair creation operator
\begin{equation}
f^+_T(q)=\sum_{u=0}^T b_T(u,q) c^+_{T-u} c^+_u
\end{equation}
Similarly the sum on $s$ yields the hermitian conjugate
$f^{}_T(q)$. Requiring normalization of the state
$f^+_T(q)|0\rangle$, some elementary algebra yields:
\FL
\begin{equation}
b_T(u,q)=\sqrt{\frac{{T \choose u}}{{T \choose q}
2^{T+1}}}\left[\frac{\partial_z^q}{q!}(1+z)^{T-u}
(1-z)^u\right]_{z=0}
\label{btuq}
\end{equation}
and
\begin{equation}
V_T(q)=\frac{I_Z(T,q)\cdot I_\xi(q)}{2\pi^2(T-q)!q!}
\label{vtq}
\end{equation}
An orthogonality relation can be easily derived from
equation (\ref{btuq}):
\begin{equation}
2\sum_{u=0}^T b_T(u,q)b_T(u,p)=\delta_{q,p}\label{btuq3}
\end{equation}
since $b_T(u,q)=b_T(q,u)$ an analogous relation holds
for sums on $q$. This relation is of frequent use when
$q$ runs only on even (odd) values. In this case the
sum can be extended to all $q$ by inserting
$(1\pm(-1)^q)/2$ in the sum. This yields
\begin{equation}
4\sum_q b_T(u,q)b_T(s,q)=\delta_{u,s}\pm\delta_{u,T-s}
\label{btuqs}
\end{equation}
the upper (lower) sign refers to even (odd) $q$.

In the limit $T\rightarrow\infty$ the coefficients $b_T(u,q)$
are simply related to Hermite orthonormal functions
$\phi_q(x)=\pi^{-1/4}(q!2^q)^{-1/2}H_q(x)e^{-x^2/2}$
($H_q(x)$ are Hermite polynomials):
\begin{equation}
b_T\left(\hbox{$\sqrt{2/T}x+\sqrt{T/2}$},q\right)=
(2T)^{-\frac{1}{4}}
\left[\phi_q(x)+O\left(\frac{q}{T}\right)\right].
\label{phiq}
\end{equation}

Note that $b_T(T-u,q)=(-1)^q b_T(u,q)$ so that, using the
commutation relations of the operators $c_u$ and
$c_u^+$, we realize that for fermions $f_T(q)\equiv 0$
for $q$ even. Similarly, as a consequence of statistics,
only even values of $q$ occur if bose particles are
concerned. This is confirmed by the symmetry of the
first quantized wave function $\psi_{T,q}(Z,\xi)$ of
the pair ($T,q$) under the exchange $\xi\rightarrow-\xi$:
\begin{eqnarray*}
\psi_{T,q}(Z,\xi)&=&\sqrt{2}\sum_{u=0}^T b_T(u,q)
\phi_u(Z+\xi)\phi_{T-u}(Z-\xi)=\\ &=&\frac{Z^{T-q}\xi^q}
{2\pi \sqrt{(T-q)!q!}}e^{-|Z|^2/2-|\xi |^2/2}
\end{eqnarray*}
This equation explicitly shows that the pair quantum
numbers $q$ and $T$ represent the RAM and the TAM
of the pair respectively.
In a classical picture the pair's center of mass
revolves round the center of the disk at a radius
$\sqrt{\langle Z^2\rangle}=\sqrt{T-q+1}$ while the two
particles rotate around the center of mass $Z$ on a
circumference of radius $\sqrt{\langle \xi^2\rangle}=
\sqrt{q+1}$. The average distance from the center of
the disk of one of the two electrons is $\sqrt{\langle
Z^2+\xi^2\rangle}=\sqrt{T+2}$. This consideration is
useful to identify the degrees of freedom relevant for
the behaviour of the system in the bulk an at the boundary.
A generalization to higher Landau levels of the expansion
in pair wave functions in RAM and TAM
was used by A.H.MacDonald et al.\cite{mcdon}.

Let us now turn to the discussion of equation (\ref{vtq}).
In the case $w(Z)=1$, $I_Z(T,q)=\pi(T-q)!$ and
$V_T(q)=\epsilon_q$ is independent
of $T$ so that:
\begin{equation}
\epsilon_q=\frac{I_\xi(q)}{2\pi q!}=\int_0^\infty
v\left(\sqrt{x}\right)\frac{x^q}{q!}e^{-x}dx
\label{eq}
\end{equation}
Note in particular that $\epsilon_q\sim
v\left(\sqrt{q}\right)$ for $q\rightarrow\infty$.
The decomposition of pair operators in components
of different RAM has already
been introduced by Haldane~\cite{pseudi} for the
spherical geometry. In the disk geometry
the same decomposition was studied by Trugman and Kivelson
\cite{trug} where a short range pair potential was
expanded in powers of its range.
An inversion formula for $v(\xi)$ as a function of
$\epsilon_q$ is obtained by multiplying equation
(\ref{eq}) by $(-1)^q{n\choose q}$ and summing over $q$:
\begin{equation}
v(r)=\sum_{n=0}^\infty\left[2\sum_{q=0}^n (-1)^q
{n\choose q}\epsilon_q\right]L_n(r^2).
\label{vdir}
\end{equation}
where $L_n(x)$ is the normalized Laguerre
polynomial of order $n$.

Examples of two body operators which can be expressed in
the form (\ref{vop}) are:
\begin{description}
\item[  i)] the Coulomb interaction
${\cal V}(z_1,z_2)=1/|z_1-z_2|$:
\begin{equation}
V_T(q)=\epsilon^c_q=\frac{\sqrt{\pi}}{2}
\frac{(2q)!}{(q!2^q)^2}
\end{equation}
note that for $q\gg 1$ $\epsilon_q\propto 1/\sqrt{q}$ as
in the classical case.
\item[ ii)] The hard core (HC) potential
${\cal V}(z_1,z_2)=2\pi\nabla^2 \delta (z_1-z_2)$:
\[V_T(q)=\epsilon^h_q=\delta_{q,1}.\]
\item[iii)] The pair correlation function $g(r)$
corresponds to ${\cal V}(z_1,z_2)=\delta(r-z_1+z_2)$:
\begin{equation}
V_T(q)=g_q(r)=\frac{r^{2q}}{q!4^q} e^{-r^2/4}.
\label{gr}
\end{equation}
\end{description}

Once the number of pairs with a definite RAM
and TAM $N_T(q)=\langle f^+_T(q)
f^{}_T(q)\rangle$ is known on a given state for all
$T$ and $q$ we are in a position to evaluate all
relevant quantities.

The operators $f^+_T(q)$ do not represent
true bosonic particles. In fact the commutation rules
for $f^+_T(q)$ has a residual term which contains a
density excitation:
\begin{eqnarray}
\left[f^{}_T(q),f^+_R(p)\right]&=&\delta_{T,R}
\delta_{q,p}-\\ \nonumber
&-&4\sum_u b_T(u,q)b_R(u,p)c^+_{R-u}c^{}_{T-u}
\label{commut}
\end{eqnarray}
The occurrence of such weird commutation relations
makes the description of the Hilbert space in terms
of pairs problematic. While
only $n/2$ pairs are necessary to build a state,
this contains much more pairs.
Indeed it is easy to check that
\begin{equation}
\sum_{T=1}^\infty\sum_q \hat{N}_T(q)=\sum_{T=1}^\infty
\sum_{u=0}^T c^+_u c^{}_u c^+_{T-u}c^{}_{T-u}=
\frac{1}{2}n(n-1).
\end{equation}
This trivial sum rule simply states that $n(n-1)/2$ pairs
can be made out of $n$ particles.

As an example of the pair's momentum distribution
function it is easy to check that for the $\nu=1$
state (all orbitals occupied up to $n-1$) all pair
states are occupied for $T<n$, i.e.
$N_T(q)=1$ $\forall q$, while
\begin{equation}
N_T(q)=2\sum_{u=T-n+1}^{n-1}b^2_T(u,q)
\hbox{~~for $T\geq n$}
\label{nt1}
\end{equation}
Figure \ref{fig1} shows the
distribution of RAM
$N(q)=\sum_T N_T(q)$ for $\nu=1/3$ and 6 electrons.
Full dots refers to the true ground state of the Coulomb
interaction while open squares to the state made of three
pair creation operators with $T=q=15$ acting on the
vacuum. The latter state contains a lot of pairs
with small RAM, including
angular momentum one. The true ground state is
instead characterized by a minimal number of pairs
with the smallest value of $q$, since these give
the largest contribution to the repulsive energy.
In particular the Laughlin state (open dots) does
not contain any pair with $q=1$ so that it is the
exact ground state for the HC interaction
\cite{girv,trug}. The dependence of $N(q)$ on $q$
has been analyzed numerically for system of $n\leq 8$
electrons. A fairly good scaling of the form
$N(q)=A_q n-B_q$ has been observed with $A_3=0.7965$,
$A_5=0.5856$, $A_7=0.4659$ and $B_q\simeq 1$ for all
$q$.

Equation (\ref{gr}) expresses a connection between the
coefficients of different powers of $r$ in $g(r)$ and
the number of pairs with a definite RAM.
These coefficients have been studied
extensively by Yoshioka~\cite{yosh}, in rectangular
geometry (Landau gauge), who has found that the
coefficient of $r^2$ and of $r^4$ (which is non zero
in this gauge) decreases by decreasing $\nu$ and
vanishes for $\nu\geq 1/3$. The same happens to the
coefficient of $r^6$ and $r^8$ for $\nu\simeq 1/5$.
A similar result was derived by Trugman and Kivelson
\cite{trug} in the symmetric gauge. In the pair
language the quantization in the FQHE occurs as a
consequence of the successive elimination of all the
pairs with the smallest RAM.

\section{Edge waves in the quantum Hall effect}

We consider, in this section, the spectrum of low lying
excitations on incompressible quantum Hall states in
disk geometry.
These excitations have been called edge waves (EW) because
they involve density fluctuations of the two dimensional
electron gas at the boundary of the system. A review on
the subject can be found in a recent paper of
X.G.Wen~\cite{wen} where a general theory for edge
excitations is discussed. The starting point of Wen's
theory is a classical hydrodynamical approach where
coordinates and canonical momenta yield, upon
quantization, the creation operators of the edge modes.
The outcome of this approach is a free phonon theory.
This is obvious if only the one body confining
potential, coming from electron-background interaction,
is considered. This picture however holds even
in the presence of the e$^-$- e$^-$ interaction.
A strong evidence of this has been given by M.Stone et
al.~\cite{stone} which have analyzed the energy spectrum
due to the pair interaction using exact diagonalization
for systems of up to $400$ particles near $\nu=1$.
The ground state $|\psi_0(n)\rangle$ for $\nu=1$ and $n$
electrons is in first quantization (apart
from the gaussian factors) the Vandermonde determinant
$\Psi_o(z_1,\ldots,z_n)=\prod_{i<j} (z_i-z_j)$
whose total angular momentum is $L_o=\frac{1}{2}n(n-1)$.
In the sector of total angular momentum $L=L_o+M$
the energy spectrum reduces, with excellent accuracy
\cite{stone}, to
\begin{equation}
E=E_o-\sum_k n_k \Omega_k
\label{edgws}
\end{equation}
where $n_k$ are (integer) bosonic occupation numbers
such that $\sum_k n_k k=M$ and $\Omega_k\geq 0$ are
{\em single particle} energies.
Note that, since only the interaction potential is
considered, the contribution to the EW spectrum is
negative as a consequence of a loss of repulsive energy.
If strictly only the LLL orbitals are considered, the
Hilbert space of the system with $L=L_o+M$ is spanned,
in first quantization, by the wave functions obtained
by multiplying $\Psi_o$ by symmetric polynomials
of degree $M$. These in turn can be expressed in a unique
way~\cite{stone2} as polynomials of {\em power sums}
${\cal S}_k(z_1,\ldots,z_n)=\sum_{i=1}^n z_i^k$.
In their work M.Stone et al. conjecture that these
polynomials correspond to the bosonic creation operators
of edge modes.

In this section the problem is reformulated
in the language of second quantization and the second
quantized counterpart of ${\cal S}_k$ are shown to
describe edge excitations as free bosons in the limit
$n\to\infty$. Next the dispersion relation due to the
interaction potential is discussed also relying on a
Hartree approximation.
We shall, as in the work of Stone et al., assume no
confining potential so that the
Hamiltonian consists only of the e$^-$- e$^-$ interaction
energy. The competition between the confining potential
and the pair interaction in the EW spectrum
is discussed elsewhere\cite{john}.
While bulk excitation involve the small $q$ part $N_T(q)$
we expect that edge excitations depend on the large $T$
behaviour of this distribution, since electrons on the edge
of the quantum dot belong to pairs with $T\sim 2n$. On
the scale of the total angular momentum $L=L_o+M$, a simple
hydrodynamical argument \cite{wen} shows that edge excitations
involve changes $\Delta L=M\sim \sqrt{n}$ (contrary to quasi
particle excitations for which $\Delta L=M\sim n$).
In the thermodynamic limit typical values of $M$ and $k$
are of the order $\sqrt{n}$. Note also that the size of the
Hilbert space depends only on $M$ as long as $M<n$.

${\cal S}_k$ is, in second quantized form, the single
particle ladder operator
\begin{equation}
S^+_k=\sum_{m=0}^\infty \sqrt{\frac{(m+k)!}{m!}}
c^+_{m+k}c^{}_m
\end{equation}
The conjugate operators $S^-_k$ are easily defined.
In general these do not satisfy commutation relations
typical of bosonic creation and destruction operators.
However M.Stone~\cite{stone2} showed, using polynomial
algebra, that, in the limit $n\rightarrow\infty$ and for
$\nu \simeq 1$, they do form a bosonic set of creation
operators in the sense that the overlap between states
with different occupation numbers (i.e. with different
combinations of $S^+_k$) vanishes. The same can be shown
to hold in the framework of second quantization for the
operators $S^+_k$. If
$|k\rangle=A_k(n)S^+_k|\psi_0(n)\rangle$ and
$|k,j\rangle=A_{k,j}(n)S^+_kS^+_j|\psi_0(n)\rangle$
are normalized states, a straightforward calculation
shows that $A_k(n)\propto n^{-k/2}$ and $\langle k,j|k+j
\rangle\propto n^{-1}$. The general $m$ {\em bosons} state,
in the sector $L=L_o+M$, is
\begin{equation}
|k_1,\ldots,k_m\rangle=A_{k_1,\ldots,k_m}(n)S^+_{k_1},
\ldots,S^+_{k_m}|\psi_0(n)\rangle
\label{base}
\end{equation}
where $\sum_i k_i=M$ and $A_{k_1,\ldots,k_m}(n)\propto
n^{-M/2}$. The overlap between states with different
occupation numbers vanishes at least as $n^{-1}$.
The state $|\psi_0(n)\rangle$ is the vacuum state
for the {\em bosons} created by $S^+_k$ since it is
annihilated by $S^-_k$ for any $k$.

If these were free bosons we would have
$\left[\hat{H},S^+_k\right]=-\Omega_k S^+_k$.
This commutator, if the Hamiltonian $\hat{H}$ has the
form (\ref{vop}), can be written as:
\begin{eqnarray*}
\FL
\left[\hat{H},S^+_k\right]&=&\sum_{T=1}^\infty\sum_q
\left\{V_{T+k}(q)f^+_{T+k}(q) \left[f^{}_{T+k}(q),
S^+_k\right]+\right.\\ &+& \left. V_T(q)\left[f^+_T(q),
S^+_k\right]f^{}_T(q)\right\}
\end{eqnarray*}
If $C^{}_{k,l}(T,q)$ is defined such that
\begin{equation}
\left[f^{}_{T+k}(q),S^+_k\right] =\sum_l
C^{}_{k,l}(T,q) f^{}_T(q-l).\label{comm1}
\end{equation}
then~\cite{nota} $\left[f^+_T(q),S^+_k
\right]=-\sum_l C^{}_{k,l}(T,q) f^+_{T+k}(q+l)$.
The coefficients $C_{k,l}(T,q)$ are easily evaluated with
the help of equation (\ref{btuq3}):
\widetext
\begin{equation}
C^{}_{k,l}(T,q)=4\sum_{u=0}^T b_T(u,q)
\sqrt{\frac{(u+k)!}{u!}}b_{T+k}(u+k,q+l)
=2(-1)^l{k \choose l}
\sqrt{\frac{(T+k-q-l)!(q+l)!}{2^k(T-q)!q!}}\label{ckltq}
\end{equation}
clearly $l\leq k$ otherwise $C^{}_{k,l}(T,q)\equiv0$.
Since only even values of $l$ occur, we will neglect
factors $(-1)^l$ and implicitly assume, from now on,
that $\sum_l$ runs only on even values.
The final outcome reads:
\begin{equation}
\left[\hat{H},S^+_k\right]=\sum_{T,q,l}C^{}_{k,l}(T,q)
\left[V_{T+k}(q+l)-V_T(q)\right] f^+_{T+k}(q+l)f^{}_T(q)
\label{hs2}
\end{equation}
\narrowtext
that is: $S^+_k$ promotes a pair in the state $(T,q)$
to states of larger TAM and RAM
$(T+k,q+l)$. $C_{k,l}(T,q)$ describes how $k$
additional units of angular momentum are distributed
among the relative and the center of mass motion of the
pair $f^+_{T+k}(q+l)$. The energy, for any such
transition, changes by an amount $V_{T+k}(q+l)-V_T(q)$.
Since $C^{}_{k=1,l}(T,q)\propto\delta_{l,0}$,
if $V_T(q)=\epsilon_q$ does not depend on $T$,
we find $\left[\hat{H},S^+_1\right]\equiv 0$.
This is a known result \cite{trug,stone,girv}:
$S^+_1$ create zero energy excitations because
these concern translations of the center of mass.
Another consequence of equation (\ref{hs2}) is that
in the case of the HC potential the commutator
turns out to be proportional to $f^{+}_{T+k}(1+l)f^{}_T(1)$
so that it annihilates the ground state for $\nu\leq 1/3$
thus yielding $\Omega_k\equiv 0$ for all $k$. Note finally
that the EW spectrum is unaffected by a constant shift
of the pair potential $V_T(q)\to V_0+V_T(q)$.

The operator form of equation (\ref{hs2}) strongly
resembles the action of a ladder operator in the relative
and center of mass coordinates. Indeed the same procedure
of section I leads, for the first quantized operator
${\cal A}_{k,l}(z_1,z_2)=Z^{k-l}\xi^l$, to the following
expression:
\FL
\[A^+_{k,l}=\frac{1}{2}\sum_{T,q}
\sqrt{\frac{(T+k-q-l)!(q+l)!}{(T-q)!q!}}
f^+_{T+k}(q+l)f^{}_T(q)\]
This is easily expressed in $S^+_k$ operators by
expanding $Z^{k-l}\xi^l$ in the individual particle
coordinates $z_1$ and $z_2$. The second quantization
procedure applied to the resulting expression reads
\begin{eqnarray}
A^+_{k,l}&=&2^{-\frac{k}{2}}\sum_{g=0}^k
\left[\frac{\partial_x^g}{g!}(1+x)^{k-l}
(1-x)^l\right]_{x=0}S^+_{k-g}S^+_g+\nonumber \\
&-& 2^{1+\frac{k}{2}}\delta_{l,0}S^+_k.
\label{Aklss}
\end{eqnarray}

Equation (\ref{hs2}) can be expressed in terms of operators
$A^+_{k,l}$ provided that $V_{T+k}(q+l)-V_T(q)$ is independent
of $T$ and $q$. Actually this is only fulfilled by
$V_T(q)=-\gamma q$ which corresponds, in real space (see
eq.(\ref{vdir})), to an harmonic potential
$v(r)=2\gamma (1-r^2)$ in the inter-particle distance $r$.
This potential has been studied\cite{parab1,parab2} in
first quantization and is a rather unphysical pair interaction
being unbounded from below as $r\to\infty$.
It has been pointed out recently\cite{parab2}
that this harmonic interaction leads to the
disappearance of the FQHE. In the present context
the stability of the FQH ground
state, that depends on the competition between the
pair interaction and the confining
potential, is assumed.
Given the harmonic potential $\epsilon_q=-\gamma q$,
equation (\ref{hs2}) is readily translated into
\begin{eqnarray*}
&~&\left[\hat{H},S^+_k\right]=-2^{2-\frac{k}{2}}\gamma\sum_l
l{k \choose l}A^+_{k,l}=\\
&=&-2\gamma\sum_{g=0}^k{k\choose g}^{\frac{1}{2}}
\left\{4\sum_l l\sqrt{{k\choose l}
2^{-k-1}}b_k(l,g)\right\}S^+_{k-g}S^+_g
\end{eqnarray*}
here the sum on $l$ runs only on even values.
The second term of (\ref{Aklss}) does not contribute (this
is true whenever $V_T(q)=\epsilon_q$).
Since $2l\sqrt{{k\choose l}2^{-k-1}}=kb_k(l,0)+
\sqrt{k}b_k(l,1)$, using (\ref{btuqs}) the sum in braces
gives $k(\delta_{g,0}+\delta_{g,k})-\sqrt{k}(\delta_{g,1}+
\delta_{g,k-1})$ and finally
\begin{equation}
\left[\hat{H},S^+_k\right]=-2\gamma k\left(nS^+_k-
S^+_1S^+_{k-1}\right)
\label{up}
\end{equation}
Note that this is an exact result ($S^+_0\equiv n$).
When evaluating the Hamiltonian for $L=L_o+M$ using the
basis set (\ref{base})
the first term will contribute to diagonal elements
while the second to off-diagonal ones. The larger off
diagonal element is the one between states
$|\ldots,k,\ldots\rangle$ and $|\ldots,k-1,1,\ldots
\rangle$. The magnitude of the latter will be of order
$n^{-1}$ with respect to diagonal elements because of
the explicit factor of $n$ in equation (\ref{up}).
Other off diagonal elements, being proportional to the
overlap of states with different occupation numbers,
will be at least of order $n^{-2}$ with respect to
diagonal elements.

In the language of standard perturbation theory the
first term of equation (\ref{up}) may be regarded as
coming from the unperturbed hamiltonian $H_o$, the second
from a perturbation $\lambda V$. The validity of
perturbation theory depends on the ratio between $\lambda$
and the separation $\Delta E$ between unperturbed
eigen-energies (diagonal elements). Equation (\ref{up})
immediately yields $\lambda/\Delta E \simeq k/n$.
First order perturbation theory is then exact as $n\to \infty$
even for physical wave vectors for which
$k\propto\sqrt{n}$: $S^+_k$ create
free edge excitations with $\Omega_k=2n\gamma k$. The
generalization to more physical pair potentials insists on
this same argument. Before turning to the general case
it is useful to remark that for $k=1$ the correct result
$\Omega_{k=1}=0$ is recovered and that the resulting
EW dispersion is linear. The spectrum is then
degenerate since $\Omega_{k+m}=\Omega_k+\Omega_m$.
If $\gamma>0$ the boson energy is an increasing
function of $k$. In general if $\epsilon_q>0$ is
monotonically decreasing $\Omega_k$ is monotonically
increasing in $k$.

An approximation scheme has to be introduced at this
point to deal with more physical interaction potentials.
The approximation essentially consists in taking
out of the sum on $T$ and $q$ an effective value
$\gamma_{k,l}(n)$ of $V_T(q)-V_{T+k}(q+l)$.
The $T$ dependence of $V_T(q)$,
together with the second term of equation (\ref{Aklss}),
will be dropped from now on.
In practice the effective value of
$\gamma_{k,l}(n)$ may be evaluated by evaluating the
$l^{\rm th}$ term of equation (\ref{hs2}) on the ground state:
\widetext
\begin{equation}
\gamma_{k,l}(n)=\frac{1}{Z_{k,l}}\langle k|\sum_{T,q}
C_{k,l}(T,q)(\epsilon_q-\epsilon_{q+l})f^+_{T+k}(q+l)
f^{}_T(q)|\psi_0(n)\rangle
=\sum_q P_{k,l}(q,n)(\epsilon_q-\epsilon_{q+l})
\label{gkln}
\end{equation}
where $Z_{k,l}$ is such that $\sum_qP_{k,l}(q,n)=1$.
The evaluation of $P_{k,l}(q,n)$ is a tedious task of
algebra that is omitted here since it leads to a complex
and lengthy expression from which it is hard to extract
the interesting properties\cite{nota2}.

With this approximation equation (\ref{hs2}) can be
expanded in $A^+_{k,l}$ operators and finally in $S^+_g$
operators
\begin{eqnarray}
\left[\hat{H},S^+_k\right]&=&-2\sum_l{k \choose l}
\gamma_{k,l}(n)A^+_{k,l}=
\frac{-1}{2n}\sum_{g=0}^k\Omega_{k,g}(n) S^+_{k-g}S^+_g
\label{ult}\\
\hbox{where~}&~&
\Omega_{k,g}(n)=4n\sum_l{k \choose l}\frac{\gamma_{k,l}(n)}
{2^k}\left[\frac{\partial_x^g}{g!}(1+x)^{k-l}
(1-x)^l\right]_{x=0}.
\label{wkgn}
\end{eqnarray}
\narrowtext
The same considerations following equation (\ref{up})
show that only the $g=0,k$ terms are dominant as
$n\to\infty$. Moreover in this limit also the approximation
$\epsilon_q-\epsilon_{q+l}\simeq \gamma_{k,l}(n)$ becomes
exact for smooth monotonic potentials. Here smooth means
that $\epsilon_q-\epsilon_{q+l}$ introduces a negligible
dependence on $q$ compared to the dependence
of $C_{k,l}(T,q)$ on the same variable. If the scale
of $q$ values is proportional to $n$ this condition is
satisfied by very general potentials. The latter condition
is verified if $P_{k,l}(q,n)$ depends on $q$ only through
the ratio $q/n$:
\begin{equation}
P_{k,l}(q,n)=\frac{1}{n}\mu_{k,l}(q/n)
\label{mukl}
\end{equation}
The reason for this is that the dependence on $n$ comes into
$P_{k,l}(q,n)$ through the pair distribution functions
$N_T(q)$ which has this property. Also one can argue that
the typical value of $q$ for a pair of electrons on the
edge of the sample is of order $n$ since the inter-pair
distance, $\sqrt{q}$, should be of the order of the disk
radius $R(n)\sim\sqrt{n}$.
Equation (\ref{mukl}) has been verified numerically.
Figure \ref{fig2} shows a very good collapse of
$nP_{k,l}(xn,n)$ for $n=30,40$ and $k=8$, $l=2,4,6,8$.

The condition on monotonicity is relevant because
otherwise $\epsilon_{q+l}-\epsilon_q$ would change
sign for some $q$. The operator structure of
$\left[\hat{H},S^+_k\right]$ could in this
case be different from that of $S^+_{k-g}S^+_g$.
Another source of troubles if $\epsilon_q$ is not
monotonic is that the sign of $\gamma_{k,l}(n)$
may change for different $l$ and this may eventually
cause cancellation in the $g=0,k$ term.
Instead, for monotonic potentials $\epsilon_{q+l}-
\epsilon_q$, and thus $\gamma_{k,l}(n)$, has always the
same sign, so that the $g=0$ and $g=k$ terms in
(\ref{wkgn}) are always at least of the same order of
magnitude of the other ones. For example table \ref{tab1}
lists the overlap between the state $\langle k|$ and the
ground state in the sector $L=L_o+k$ and $n=30$,
for the Coulomb, the HC potential and for
the potential $\epsilon^x_q=q/(q^2+16)$. The latter
is not monotonic and has a maximum at $q=4$.
We see that in the former cases the overlap is very
close to unity, while for $\epsilon^x_q$ this is
not true for $k>5$. At $k=6$ the state $\langle k|$
is between the ground state and the first excited
state while for $k=7$ it is very close to the second
excited state. Another evidence of the correctness of
the approximation is that the overlap depends weakly
on the potential.

We can safely conclude that $S^+_k$ are the
creation operators of the edge modes
\[\left[\hat{H},S^+_k\right]=-\Omega_{k,0}(n)S^+_k
\hbox{~~as $n\to\infty$}\]
where $\Omega_{k,0}=\Omega_{k,k}$ is given by (\ref{wkgn}).
In the case $k\gg 1$, a rough estimate of $\Omega_{k,0}(n)$
is obtained observing that in (\ref{wkgn}) for $g=0,k$ a
binomial average is performed of $\gamma_{k,l}(n)$ and then
$\Omega_{k,0}\simeq 2n\gamma_{k,k/2}(n)$.

Let us analyze in more detail the EW spectrum
resulting from a general potential.
Of particular interest is the behaviour of $\Omega_{k,0}(n)$
for a fixed $k$ as $n\rightarrow \infty$. This is related
to the behaviour of the dispersion relation of EW
in the long wavelength limit. In fact the physical wave
vector $\kappa$ is proportional\cite{wen} to $k/\sqrt{n}$.
The hydrodynamical picture\cite{wen} assumes a dispersion
relation linear in $\kappa$ for $\kappa\to 0$.
The velocity $c$ of EW is given by
\begin{equation}
c=\lim_{\kappa\to 0}\frac{\Omega_{\kappa\sqrt{n},0}}{\kappa}=
\lim_{n\to \infty}\frac{\sqrt{n}\Omega_{k,0}(n)}{k}
\label{sound}
\end{equation}
As long as the potential $\epsilon_q$ is
monotonic, $\gamma_{k,l}(n)$ will always contain
a linear term in $l$ so that the linear term in $k$
of $\Omega_{k,0}$ will always be present. The
evaluation of the explicit $k$ dependence of
$\Omega_{k,0}(n)$ is complicated by the dependence
introduced by $P_{k,l}(q,n)$ which is difficult to
analyse.
The scaling property of $P_{k,l}(q,n)$, equation
(\ref{mukl}), allows however to draw
conclusion on the dependence on $n$ of
$\Omega_{k,0}(n)$. These conclusions will be
further confirmed by an Hartree approximation for
$\nu=1$ that will also give some indication
on the $k$ dependence of the dispersion relation.

The method is based on the separation of the
dependence on $l$ and $q$ of $\epsilon_q-\epsilon_{q+l}$.
If $\epsilon_q-\epsilon_{q+l}=\alpha(l)\beta(q)$ then
$\gamma_{k,l}(n)=\alpha(l)\eta_{k,l}(n)$ where
$\eta_{k,l}(n)=\sum_qP_{k,l}(q,n)\beta(q)$. The dependence
on $n$ of $\eta_{k,l}(n)$ can be extracted using
equation (\ref{mukl}), i.e.
\begin{equation}
\eta_{k,l}(n)=\int\,dx\mu_{k,l}(x)\beta(nx).
\label{etakl}
\end{equation}

In the case of a finite range potential of the form
$\epsilon_q=e^{-q/q_o}$ we may take $\alpha(l)=1-
e^{-l/q_0}$ and $\beta(q)=e^{-q/q_o}$
that in equation (\ref{etakl}) means that the behaviour
of $\eta_{k,l}(n)$ as $n\to\infty$ is related to the
behaviour of $\mu_{k,l}(x)$ as $x\to 0$ as could be expected
since only pairs with small RAM contribute to the energy.
Also note that for $l\simeq k/2\propto \sqrt{n}$ all the
dependence on $k$ and $n$ is in $\eta_{k,l}(n)$ since
$\alpha(k/2)\simeq 1$. However the analysis of
$\mu_{k,l}(x)$ as $x\to 0$ is a very difficult task.
We will circumvent this difficulty introducing the Hartree
approximation in the following. A point worth of mention
here is that the behaviour in $n$ of $\Omega_{k,0}(n)$ for
short range potentials depends on the properties of the
ground state, i.e. on $\mu_{k,l}(x)$. No assumption
on the form of $\mu_{k,l}(x)$ is instead necessary to
extract the $n$ dependence of $\Omega_{k,0}(n)$
in the case of a long range potential of the form
$\epsilon_q=q^{-a}$ (here $a=1/2$ would correspond to
the Coulomb potential). The separation of the $l$
dependence from the $q$ dependence is possible
in this case using power expansion. The term $l^m$ will
have a coefficient $\beta_m(q)\propto
q^{-a-m}$ that in equation (\ref{etakl}) will give a
term of the order of $n^{-a-m}$. In the thermodynamic
limit the $m=1$ term is dominant and all the others
can be neglected even when $l\propto\sqrt{n}$.
The situation is then very similar to that
of the harmonic potential since $\epsilon_q-
\epsilon_{q+l}$ can be replaced by $l\sum_q P_{k,l}(q,n)
aq^{-a-1}$.
The resulting spectrum $\Omega_{k,0}$ will be linear in $k$
in a first approximation. However, due to the additional
dependence on $l$ and $k$ introduced by $P_{k,l}(q,n)$,
the EW frequency could contain also higher powers
of $k$. The coefficient of the linear term in
$\Omega_{k,0}(n)$ will be of the order
of $n^{-a}$ that in equation (\ref{sound})
yields a sound velocity $c\sim n^{1/2-a}$ that vanishes
for $a>1/2$. In real space this means that for potential
vanishing faster than $1/r$, that is faster than the
Coulomb potential, the contribution of the
interaction energy to the EW velocity is zero.

Let us now turn to the Hartree approximation to the
EW frequency. The derivation of this
approximation is presented in the appendix. Here we
briefly comment on its nature before turning to the
discussion of the results.
The final formula can also be obtained from
equation (\ref{hs2}) by replacing the pair of
operators $f^+_{T+k}(q+l)f^{}_T(q)$
with their commutator that has a form similar to $S^+_k$
(see (\ref{commut})) of a density excitation.
The justification for this is that the neglected term,
$f^{}_T(q)f^+_{T+k}(q+l)$, tries to create pairs with
TAM $T+k$ on the filled Landau level and this is of
course not possible if $T+k\le 2(n-1)$.
Since the terms with $T>2(n-1)$ of equation (\ref{hs2})
vanish on the $\nu=1$ state (since $f^{}_T(q)|\psi_0(n)
\rangle\equiv 0$) the neglected exchange term acts only on
the "Fermi surface". Apart from this, the derivation
also makes use of equation (\ref{phiq}) and then
the condition $q\ll n$ is assumed.
The final result reads
\begin{equation}
\Omega^{{}^H}_k(n)=\sum_{j=1}^\infty
\left(\frac{k}{n}\right)^j
\sum_q\epsilon_q\Gamma^{(j)}(q/n)
\label{omegah}
\end{equation}
The first two terms $j=1$ and $2$ terms
were worked out explicitly and their expansion in powers
of $\sqrt{q/n}$ were found to contain only odd powers.
The dependence on $q$ through the ratio $q/n$ stresses
once again that edge wave excitations involve pairs of
RAM of order $n$.
Note that, for $k\propto\sqrt{n}$, the linear term in
$k$ dominates on all the others and this suggests that
the asymptotic spectrum is linear. This means that the
harmonic approximation for the Hartree potential is
a good one\cite{nota3}.

Since $\Gamma^{(1)}(y)\propto \sqrt{y}+O(y^{3/2})$, it is
straightforward that for a finite range potential
$\Omega^{{}^H}_k(n)\sim kn^{-3/2}$. This result coincides
with the power law behaviour found by Stone et.al.\cite{stone}
numerically. If combined with equation (\ref{etakl}), it
suggests that $\mu_{k,l}(x)\sim x^{3/2}$ for $x\ll 1$.
Note that for a short range interaction also the possibility
of a finite dispersion relation with a long wavelength
behaviour of the type $(k/\sqrt{n})^3$ is ruled out by
equation (\ref{omegah}). The above mentioned results for
a long range potential are nicely recovered together with
the numerical result\cite{stone} $\Omega_k(n)\sim k
n^{-1/2}$ for the Coulomb potential. Moreover equation
(\ref{omegah}) provides a further indication for the
asymptotic linearity of the dispersion relation for long
range potentials.
Finally the coefficient of the $k^2$ term, that removes
the degeneration of the linear term, were found to be
positive so that, as conjectured by Stone et al.\cite{stone},
the ground state in the sector $L=L_o+M$ is actually
$S^+_M|\psi_0(n)\rangle$.

In summary it has been shown that for monotonic potentials
edge wave excitations near $\nu=1$ are free bosons as
$n\to\infty$. The creation operators of edge excitations are
$S^+_k$. This result is in some sense a consequence of the
completeness of the basis set created by $S^+_k$. Every
operator that raises the total angular momentum of the system
can be expressed as a sum of combinations of $S^+_k$. This
has been explicitly done for the commutator $[\hat{H},
S^+_k]$ which turn out to have a dominant contribution,
as $n\to\infty$, proportional to $S^+_k$.

The behaviour of the EW spectrum has been discussed
for general classes of potentials. In particular the
long wavelength limit, that is related to the $n\to\infty$
limit for $k\sim\sqrt{n}$, has been explored in a general
way using the fact that pairs with a RAM of order $n$
are involved in edge excitations. This statement, that
refers to the difference in the RAM distribution of the
excited state with respect to that of the ground state,
is also displayed in the results of the Hartree
approximation, where again the dependence on the RAM
$q$ comes through the ratio $q/n$.
The fact that macroscopic quantum numbers are
involved in edge excitations supports the validity of
the classical hydrodynamical picture\cite{wen}.
The special role played by the harmonic $\epsilon_q=
-\gamma q$ interaction and the result, from the
Hartree approximation, that the asymptotic dispersion
is linear in $k$, is also reminiscent of
a classical elastic response. In some sense this
potential comes out naturally in dealing with edge
excitations for any pair interaction $\epsilon_q$;
the effective $\gamma(n)$ being some
average of $\partial_q\epsilon_q$. The reason why
this comes out is essentially the same for which the
single particle confining potential, coming from
Coulomb interaction with nuclei, is usually modeled
by a harmonic one\cite{john,parab1,parab2,mcdonj,nota3}.

A final consideration concerns the extension of these results
to $\nu=1/m$. Note that equation (\ref{up}) for the
potential $\epsilon_q=-\gamma q$ is an exact result for any
$\nu$. This formula, as equation (\ref{ult}), is a statement
about operators.

There are two basic conditions on the ground state
$|\psi_0(n)\rangle$ that have been used:
\begin{description}
\item[  i)] $S^+_k$ provide an orthonormal basis of
the Hilbert space for $L=L_o+M$. The overlap between
different states $\langle j_1,\ldots,j_m|k_1,\ldots,
k_p\rangle$ vanishes as $n^{-1}$.
\item[ ii)] $S^-_k$ annihilate the ground state
for any $k$: $S^-_k|\psi_0(n)\rangle\equiv 0$.
\end{description}
If these two conditions are satisfied also by the
Laughlin state then the whole description of EW
presented for $\nu=1$ can be applied at $\nu=1/m$.
Actually if condition {\bf i)} holds, the second one follows
by a simple argument. This is because if it were not so
also $S^-_k$ would provide low lying excitations so that
the ground state energy would not have a cusp at $\nu=1/m$.
Evidences for the validity of condition {\bf i)} for $\nu=1/m$
come from the parton construction of the Laughlin states
\cite{jain,wen,stone2} and also from the result
$\left[\hat{H},S^+_k\right]=0$ for the HC interaction.
These strongly suggest that the same picture outlined
for $\nu=1$ holds for $\nu=1/m$. This supports the
idea\cite{mcdonj} that edge states at $\nu=1/m$ are in
a 1 to 1 correspondence to edge states at $\nu=1$.
Another consequence of this is that, while the asymptotic
behaviour of the EW spectrum is expected to change
for finite range potentials, as it does change for the
HC potential, going from $\nu=1$ to $\nu=1/m$
(here $\Omega_k\equiv 0$), the same
behaviour is expected for long range potentials. This
is because the power law behaviour of $\Omega_{k,0}(n)$
on $n$ depends explicitly on the structure of the ground
state only in the former case. The conclusion
that the Coulomb interaction gives contribution to
the EW velocity while potentials vanishing faster
as $r\to \infty$ do not, can then be extended to $\nu=1/m$.

\section*{Acknowledgments}

I wish to thank E.Tosatti for
frequent discussions and for proof reading of the
manuscript. I am also grateful to M.Bernasconi and
G.Baskaran for very interesting discussions and to
M.D.Johnson for drawing my attention to
ref.~\cite{stone}, for valuable discussions and for
a stimulating correspondence.

\section*{Appendix: The Hartree approximation}

The pair interaction hamiltonian can be approximated
by a single particle potential for $\nu=1$ using
$f^{+}_T(q)f^{}_T(q)=f^{}_T(q)f^{+}_T(q)-
\left[f^{}_T(q),f^+_T(q)\right]$. If only the
part of $f^{+}_T(q)f^{}_T(q)$ acting on the lowest $n-1$
orbitals is considered the first term
can be neglected. The hamiltonian then becomes
$\hat{V}^{{}^H}=\sum_{u<n} e_u(n) c^+_uc^{}_u$ where
the energy of the orbital $u$ is given by
\[e_u(n)=\sum_{T,q} \epsilon_q 2 b^2_T(u,q)\]
the sums on $T$ and $q$ run on all the values
consistent with $u<n$ and $T-u<n$ for the
above mentioned restriction.
This single particle potential yields indeed the
exact value (\ref{nt1}) of the $\nu=1$ interaction energy.
The commutator with $S^+_k$ is easily carried out
with the result
\[\left[\hat{V}^{{}^H},S^+_k\right]=-\sum_m
\left[e_m(n)-e_{m+k}(n)\right]\sqrt{\frac{(m+k)!}{m!}}
c^+_{m+k}c^{}_m.\]
This can be approximated by $-\Omega^{{}^H}_k(n)S^+_k$
by taking out of the sum the value of
$e_m(n)-e_{m+k}(n)$ for $m=n-1$ on the
"Fermi surface". In other words $\Omega^{{}^H}_k(n)$ is
the gain in energy if one electron on the outer orbital
is raised by $k$ units of angular momentum.
The result can be cast in the form
$\Omega^{{}^H}_{k}(n)=\sum_q\epsilon_q\Gamma_k(q,n)$ where
\[\Gamma_k(q,n)=2\sum_{T=q}^{2(n-1)}
\left[b^2_T(n-1,q)-b^2_{T+k}(n-1+k,q)\right].\]
This expression can be evaluated
using the variable $x=(n-1-T/2)/\sqrt{T/2}$ and
the asymptotic form of $b^{}_T(u,q)$ equation
(\ref{phiq}). The sum on $T$ can be changed into an integral
in $dx$ whose upper limit can be taken to by $\infty$
for $q\ll T\propto n$ (this is also the condition for the
validity of (\ref{phiq})).
After some algebra $\Gamma_k(q,n)$ takes the form
\[\sqrt{\frac{n}{\pi}}\int_0^\infty J(y)
\left[\phi^2_q(\sqrt{n}y)-A(y,g)
\phi^2_q\left(\sqrt{n} C(y,g)\right)\right]dy\]
where $J(y)dy$, $A(y,g)$ and $C(y,g)$ are
the expressions of $\sqrt{2/(nT)}dT$, $\sqrt{T/(T+k)}$ and
$(n-1+k/2-T/2)/\sqrt{n(T+k)/2}$ respectively in the variables
$y=x/\sqrt{n}$ and $g=k/n$. Note that $A(y,0)=1$ and
$C(y,0)=y$. Next the integrand
is expanded in powers of $g$ and the terms containing
derivatives of $\phi_q^2(x)$ are integrated by parts.
The final result is equation (\ref{omegah}) where
\[\Gamma^{(j)}(q/n)=\int_0^\infty F_j\left(\frac{x}{\sqrt{n}}
\right) \phi^2_q(x)dx.\]
The dependence on $q/n$ of the above expression comes
from the fact that $\phi^2_q(x)$ falls off rapidly for
$x>\sqrt{q}$. The cases $j=1$ and $2$ were
worked out explicitly with the result $F_1(y)=y-
\frac{3}{8}y^3+\frac{15}{128}y^5-\frac{35}{1024}y^7+\ldots$
and $F_2(y)=\frac{3}{4}y-\frac{15}{32}y^3+
\frac{105}{1024}y^5-\frac{315}{4096}y^7+\ldots$.

\begin{figure}
\caption{Plot of the distribution of relative angular
momentum, $N(q)=\sum_T N_T(q)$, for a $6$ fermion system
at effective filling $1/3$. $L_{tot}$ is the total
angular momentum, $m_{max}$ is the highest Landau
orbital considered. Lines are drawn only for guiding
eyes. The full dots ($\bullet$) refers to the exact
ground state, open dots ($\circ$) to the $m=3$ Laughlin
state while the squares ($\Box$) refers to the state
built with three pair creation operators of highest
relative angular momentum.}
\label{fig1}
\end{figure}
\begin{figure}
\caption{Plot of the distributions $\mu_{k,l}(q/n)=
nP_{k,l}(q,n)$ for $n=30,40$, $k=8$ and all even values
$l=2,4,6$ and $8$.}
\label{fig2}
\end{figure}
\begin{table}
\begin{tabular}{cccc}
$k$ &
$\langle k|\psi_o^c\rangle$ &
$\langle k|\psi_o^h\rangle$ &
$\langle k|\psi_o^x\rangle$\\
\tableline
2 & 0.9995 & 0.9995 & 0.9995\\
3 & 0.9967 & 0.9967 & 0.9967\\
4 & 0.9896 & 0.9905 & 0.9885\\
5 & 0.9758 & 0.9795 & 0.9611\\
6 & 0.9525 & 0.9624 & 0.6077\\
7 & 0.9168 & 0.9379 & 0.0275\\
7 &    -   &    -   & 0.9718\\
\end{tabular}
\caption{Overlap between the state $\langle k|$ and the
ground states of the Coulomb ($|\psi_o^{c}\rangle$),
of the hard core ($|\psi_o^{h}\rangle$) and of the
potential $\epsilon^x_q$ ($|\psi^x_o\rangle$)
defined in the text. In the last line
$|\psi^x_o\rangle$ is the first excited
state of the latter potential for $k=7$.}
\label{tab1}
\end{table}

\begin{thebibliography}{99}
\bibitem{nota} Here we have disregarded a term involving
 the sum over the first $k$ states of the LLL which is
 negligible.
\bibitem{nota2} The function $P_{k,l}(q,n)$ has a complex
 expression that depends not only on $N_T(q)$ but on the
 full "pair density matrix" $G_T(q,p)=\langle\psi_0(n)|
 f^+_T(q)f^{}_T(p)|\psi_o(n)\rangle$.
\bibitem{nota3} The commutator of a single particle
potential $\hat{V}_1=\sum_m E_m c^+_mc^{}_m$ with $S^+_k$
yields $ck S^+_k$ only if $E_m=c m$ that is for a
harmonic potential. For a general potential the commutator
is very close to $S^+_k$ under the same assumptions of
monotonicity and smoothness discussed in the text for
$\epsilon_q$. An effective harmonic potential can be
defined for $\nu=1$ with $c(n)\simeq partial_m E_m|_{m=n-1}$.
\end{thebibliography}
\end{document}